# Towards an MDD Based Framework for Self Adaptive IoT Applications Development

**Yousef Abuseta**

Computer Science Department, Faculty of Science, Al-Jabal Al-Gharbi University, Zintan, Libya

yousef.m.abusetta@gmail.com

## ABSTRACT

As technology and communication advances, more devices (and things) are able to connect to the Internet and talk to each other to achieve a common goal which results in the emergence of the Internet of Things (IoT) era. It is believed that IoT will bring up a limitless number of applications and business opportunities that will affect almost every aspect of our life. Research has already been conducted to investigate the challenges that obstruct the realization of IoT along with the promising solutions that pave the way for the acceptance and enabling of IoT. Among the research areas that is of a great importance to making IoT paradigm possible is the presence of a unified programming framework that masks the heterogeneity of the involved devices of the IoT platform. Such a framework guides system developers throughout the IoT application development process. In this paper, we investigate the IoT concept and its high level architecture in general and focus more on the application development aspect. We believe that IoT applications are highly dynamic in nature and thus need to be engineered with the self adaptive and autonomic concepts in mind. Therefore, our proposed IoT software development lifecycle was based on the IBM architecture blueprint for autonomic systems. To cater for the runtime dynamic and heterogeneity aspects of IoT applications, we adopt the MDD paradigm for our proposed development framework. We highlight the core requirements of a resilient development framework that accommodates the necessary concepts and processes for a successful IoT application.

## 1. INTRODUCTION

The Internet of Things (IoT) has Increasingly gained remarkable attention in industry as a way of networking and connecting different types of physical devices and forming networks of information [1]. This concept is the based on the pervasive presence of a variety of things or objects – such as Radio-Frequency Identification (RFID) tags, sensors, actuators, mobile phones, etc. – which, through unique addressing schemes, are able to interact with each other and collaborate with their neighbors to achieve common goals [2]. A definition by the International Telecommunication Union (ITU) states that the IoT is "A global infrastructure for the Information Society, enabling advanced services by interconnecting (physical and virtual) things based on, existing and evolving, interoperable information and communication technologies". Connecting these devices can be accomplished either directly through cellular technologies such as 2G, 3G and 4G or they can be connected via a gateway, forming a local area network, to get connection to the Internet. The gateway method enables forming Machine to Machine (M2M) networks via the use of various radio technologies. Popular examples of such technologies include Zigbee (based on the IEEE 802.15.4 Standard), Wi-Fi (based on the IEEE 802.11 Standard), 6LowPAN over Zigbee (IPv6 over Low Power Personal Area Networks), or Bluetooth (based on the IEEE 802.15.1) [3]. The IoT have influenced many domains such as health care, fitness, education, entertainment, social life, energy conservation, environment monitoring, home automation, and transport systems[Hindawi-4].

A large body of research has been carried out to investigate the challenges that hinder the realization of IoT as well as the promising solutions that assist in making the IoT a reality. Amongst the research areas that is of a great importance and has gained much attention to making IoT paradigm possible is the development of a unified programming framework that helps overcome the heterogeneity of the involved devices and provides a set of horizontal service components that are generic enough to accommodate various vertical applications . Such a framework guides system developers throughout the IoT application development process. In this paper, we investigate the IoT concept and its high level architecture in general and focus more on the application development aspect. We highlight the core requirements of a resilient development framework that accommodates the necessary concepts and processes for a successful IoT application.

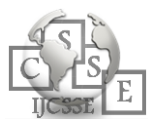

The rest of the paper is structured as follows. Section 2 serves as a background on related concepts and approaches to the work of this paper. Section 3 reviews some related work on development techniques for IoT applications. Section 4 presents the proposed IoT application development framework. The paper is concluded in section 5 with some outlined directions for future work.

## 2. BACKGROUND

### 2.1 IoT high Level Architecture

This section is dedicated to the architecture layers of IoT that have been proposed by researchers in the literature. Up until now, to the best of our knowledge, there is no an agreed upon architecture that is used by all researchers of IoT. However, there is a set of layers that is expected to be present in every proposed architecture, though it is likely to be presented with different concepts and terminology. As it consists of significantly diverse objects, IoT requires an open architecture to enable, to large extent, the interoperability among the heterogeneous systems and distributed resources [5].

Before we delve into the discussion of the IoT reference architecture, a number of important concepts, which lay foundation for the IoT paradigm, is worth presenting here. In [6], a description of the IoT domain model is introduced. Such a model is mainly based on the interaction relationship between two entities, namely the *user* and *physical entity*.

The user here is not meant to be restricted to a human but it can also be any kind of digital artifacts such as services, applications or software agents that have the interest (goal) of interacting with the physical entity. The physical entity is any identifiable object that is part of the physical environment such as humans, cars, animals, computers, electronic appliances, etc. The user role itself can be played by another physical entity in which case the Machine to Machine (M2M) interaction is established. In fact, this is the heart of the IoT paradigm in which a number of things or machines are interacting and exchanging data in order to achieve a collective goal. Interaction usually occurs indirectly via some dedicated services that would either get information about the physical entity or perform some actions on it. The latter usually changes the state of the physical entity. A virtual entity, such as an object in Object Oriented Programming, is the digital representation of a physical entity. The virtual and physical entities are usually related to each other by embedding into or placing nearby the physical entity one or more ICT devices (e.g. sensors, tags, actuators). The sensors and actuators concepts are used heavily in many paradigms such as autonomic systems, self adaptive systems and wireless sensor networks. These devices enable the technological interface to the physical entity where data can be collected and actions are applied. However, this interface is not defined directly using the sensors and actuators devices but via drivers (software components) that are able to interface with these devices and perform the read (from sensor) and command dispatching (to actuator) operations [7]. In many IoT reference architectures, the physical entity along with the ICT devices are referred to as the sensing or device layer which resides at the bottom of the architecture. The IoT application starts at this layer where physical entities send signals carrying some data to be processed, checked for violation and stored for further processing . The signals usually make use of binary proprietary protocols which vary from one physical entity to another.

As pointed out in [8], the direct communication between the physical entity and the application processing the sent data is quite difficult. It is put down to two fundamental issues: 1) the application that processes data received from devices needs to scale to each single device. 2) the security issue is compromised since the application processing the data usually uses a heavy protocol for authentication while the device usually uses a firmware that cannot be reprogrammed to have things such as passwords and certificates. To address these issues, an on field gateway is suggested. Such a gateway can be used, beside its main task, to aggregate data collected from a number of nearby devices and discover locally any possible undesirable system states. The latter helps in shielding the backend system from extra work that might affect its performance and ability to scale and manage more devices. Also, the gateway may be used as an adapter to transform a binary based protocol to a more standard protocol to be read by the other components of the system.

Consequently, a middleware layer is a crucial component of IoT architecture as it acts as a bridge between the heterogeneous devices and the enterprise applications that access them. Figure 1 shows a high level reference architecture of IoT platform.

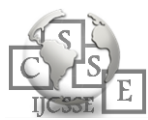

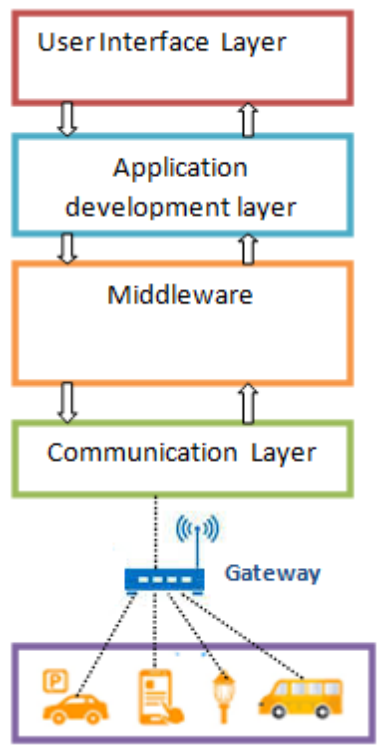

*Fig. 1. A high level reference architecture of IoT platform.*

## 2.2 Feedback control loops

Self adaptation or autonomic capabilities can be introduced to the software system either internally or externally [9]. In the internal approach, the adaptation logic (managing system) is intertwined with the core application (managed system) which may take the form of the exception handling. In this case, the adaptation engine is system dependent and thus difficult to maintain, evolve, and reuse. In contrast, in the external approach, the concerns of the adaptation logic are separated from the core application. Most of the existing approaches adopt the external approach since it enables the realization of some important software qualities such as the reusability and modifiability. The IBM architecture blueprint [10] is an example of this approach. Such an architecture is centered around the idea of the feedback control loops.

A feedback loop is a control loop where the output of the controlled system is fed back to the input. It allows therefore to adjust operations according to differences between the actual output and the desired output. In other words, feedback control loops are entities that observe the system and initiate adaption. A feedback loop typically involves four key activities: collect, analyze, decide, and act [11]. Sensors collect data from the running system and its environment which represents its current state. The collected data are then aggregated and saved for future reference to construct a model of past and current states. The data is then analyzed to infer trends and identify symptoms. The planning activity then takes place and attempts to predict the future and prepare change plan to act on the running system through a set of effectors or actuators. [12].

In IBM architecture blueprint, the managing system consists of four main activities: monitor, analyze, plan and execute. These activities share a knowledge base component which contains information about the system state as well as the policy engine that controls the system functioning. A set of sensors is used to collect the important data to the adaptation process and send them to the monitor for further processing while a set of effectors is used to apply the corrective changes stated in the plan.

## 3. RELATED WORK

Several approaches have been proposed to address the design and development of IoT applications. Here we present some well known and popular approaches to the IoT area. DiaSuite [13] offers a design language, providing high-level, declarative constructs that are dedicated to describing the application's architecture, along with the smart objects it orchestrates. HYDRA [14] is a service oriented middleware. It accommodates a set of software components used for handling many tasks required for the development of intelligent applications. A semantic interoperability is provided here using semantic web technologies. It also supports dynamic reconfiguration and self-management. IoT-A [6], has proposed a reference architecture for the development of IoT applications. This reference architecture serves as a tool for building compliant IoT architectures. it provides views and perspectives on different architectural aspects that are of concern to stakeholders of the IoT

Oracle [15], also has developed a reference architecture for the IoT platform with an emphasis on the middleware layer.

Also, Microsoft has proposed and developed a reference architecture called Azure [8] for creating and enabling IoT solutions.

## 4. Proposed IoT Application Development Framework

As we believe that the IoT applications are highly dynamic in nature, engineering such applications must be conducted with the self adaptive and autonomic properties () in mind. We believe that self adaptive system concepts should be made first class entities and thus need to be inherent from the early stages of the engineering of IoT applications. Our proposed framework therefore has adopted the IBM architecture blueprint [10] for modeling the feedback control loop that consists of the Monitor, Analyze, Plan, Execute and Knowledge base components. We also build the work presented in this paper on a previous work [16] for the engineering of autonomic systems using the Model Driven Development (MDD) technique as well as on some related design patterns [17] and proposed framework for testing and simulating Self Adaptive Systems [18].

## 4.1 Characteristics of Proposed Framework

For the success and effectiveness of the proposed framework, a number of characteristics has to be exhibited. Such characteristics are described as follows:

- Generic: it should be generic enough to be used across a variety of vertical applications and services.
- Ease of use: it should be easy to use from the point of view of system developers.
- Extensibility: it concerns with the ability of the framework to be extensible to accommodate new features and capabilities. For instance, it should be easy to introduce a new physical entity as well as new protocols that support these entities. *Customization*: it concerns with the ability of the framework to be customized and tailored for some specific systems or some organizations of feedback control loops (e.g. decentralized or hierarchical) .
- *Testability*: it concerns with the ability of the framework to be tested for some tasks and activities. Testing the process of monitoring a specific system property and taking the appropriate corrective actions is only one example.

## 4.2 Conceptual View Of Proposed Framework

This section serves as a conceptual view of the proposed framework. We adopt the modular approach where the framework is decomposed into a set of subsystems organized into a number of packages or namespaces. Each subsystem, in turn, contains a number of components (or subsystems) that cooperate to accomplish some specific tasks. Fig. 2 depicts this conceptual view of the framework which consists of the managing system, managed system and environment.

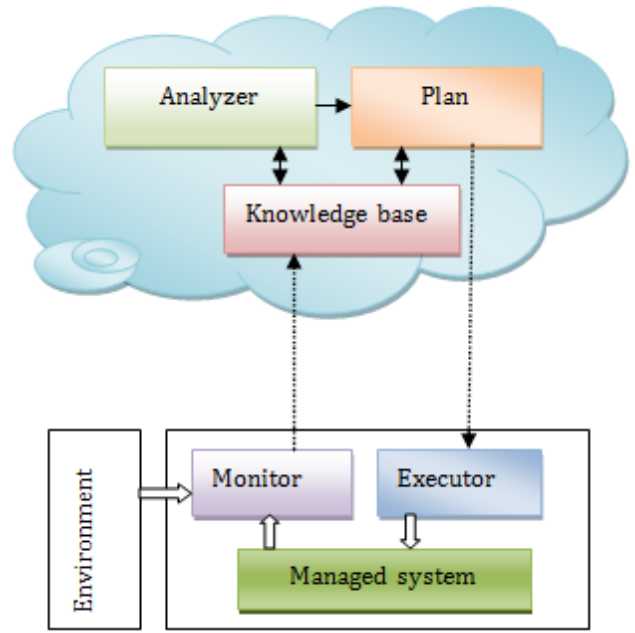

*Fig. 2. A high level reference architecture of IoT platform.*

Unlike the traditional autonomic software systems, the *managed system* here represents the physical entities or devices, as well as the virtual entities that represent them, which are of paramount importance to IoT applications. As for the *managing system*, the feedback control loop plays the management role. As discussed earlier, an IoT application is considered to be a real-time system and thus functions and responses to events in this platform must be conducted in a timely manner. To enable the backend system (probably hosted in the cloud) to be acting on time, one promising approach is to perform some of the management activities locally and avoid sending unnecessary data. This suggests modeling the monitor component as a multi function gateway nearby the managed system (devices). Also, the executor component can be positioned locally to further reduce the burden on the backend system which results in a distributed organization of the feedback control loop as shown in Fig.1. Such an organization is one of various forms of the possible interactions between the MAPE-K components which were presented and discussed in [19]. The *environment* can be defined as any external actor that affects the managed system or the adaptation decisions of the managing system in some way. Therefore, the environment property represents any contextual information that is external to the system in question and contributes to its runtime state.

## 4.3 Proposed Framework Requirements

This section serves as an analysis for the software requirements of the proposed framework. These requirements are essential to provide a flexible framework that allows developers to model and develop IoT applications and services in a seamless manner. The requirements specification process is a use case driven. To achieve the separation of concerns design principles, we consider one subsystem (e.g. managed system) at a time when defining the different use cases.

### 4.3.1 Managed system requirements

The managed system represents the system under development which is composed of the physical entities or devices that are involved in the IoT application along with the virtual entities that represent them as software components. The following is the requirement that is related to the managed system and expected to be available in the framework:

- *provide interface for IoT managed system model:* This requirement is related to providing appropriate interfaces to define and model the managed systems. Figure 3 depicts the UML use case diagram for the managed system requirement.

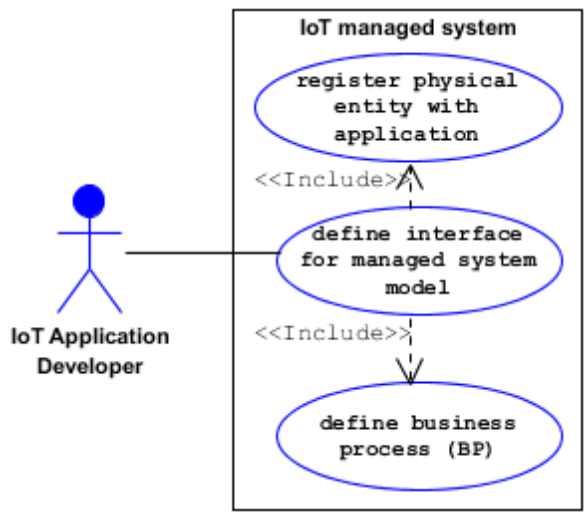

*Fig. 3. UML use case diagram for the managed system requirement.*

4.3.1.1 As stated earlier, the managed system represents the physical entities or devices along with the virtual entities that represent them as software components. Thus, one of the main requirement here is to provide a mechanism or an interface for registering and integrating the physical entity with the IoT application in question. However, the different physical entities and their virtual representations are of little value unless were modeled in the context of a set of business processes that represents the reasons behind developing such an application. Our approach starts with defining a domain where each domain consists of a set of tasks and each task is realized through the interaction of a number of services (composite). Below is a description of these concepts.

*Domain.* The domain here is the system under consideration which comprises a number of tasks. Examples of domain include the healthcare , home automation, smart metering and smart transportation.

*Task.* A task is a high level goal that has to be addressed in order to realize the overall system requirements. Each task, in turn, contains a set of services responsible for addressing and achieving that task. A task in a healthcare system is, for example, *monitor energy meter reading at home.*

*Service.* A service is an abstraction of a software or hardware entity (physical entity or device) that has a role to play in addressing the task goal. These services, later at the code generation stage, are mapped into software components such as RESTful web services, CORBA, Java, .NET, etc. A blood pressure sensor is an example of service.

*Composite.* The services of a particular task coordinate with each other to address the purpose of that task. Such coordination, which involves a set of interactions, is encapsulated in an entity called composite. A composite might contain only one service. However, a useful composite is often composed of more than one service.

### 4.3.2 Managing Systems Requirements

The managing system represents the management layer whose responsibility is to introduce autonomic capabilities to managed systems. Therefore, the requirements here are concerned with activities such as monitoring, analysis, planning and execution (in addition to the policy and symptoms definitions).

- *Monitoring system requirements:* The monitoring system should capture issues related to what, when and how to monitor. In the IoT platform, the what to monitor aspect is concerned with monitoring properties of the physical entities which are of significant importance to the managing system and keeking them within a desirable range is a key to a resilient autonomic system. The when to monitor aspect is concerned with the timing of the monitoring. Readings of interesting properties can be measured and reported to the monitor at fixed delay, in response to an event and/or on demand. How to obtain the readings of interesting properties is the concern of the how to monitor aspect. Here, we use a sensor embedded in or located nearby the physical entity to make a direct measurement of these properties.

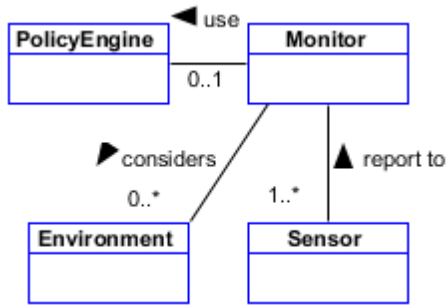

*Fig. 4. A high level structure of Monitoring system.*

Also, since the monitor is assigned the task of aggregating and filtering data collected from sensors, a local policy engine should be introduced to the monitoring system. The environment can also affect the state of the monitored

4.3.2.1 Based on the above discussion, we can list the following requirements for the monitoring activity:

- Specify device property for monitoring
- Specify monitoring mode
- Create local policy engine
- Register sensors with monitor

Figure 5 shows the UML use case diagram for the monitoring system requirements.

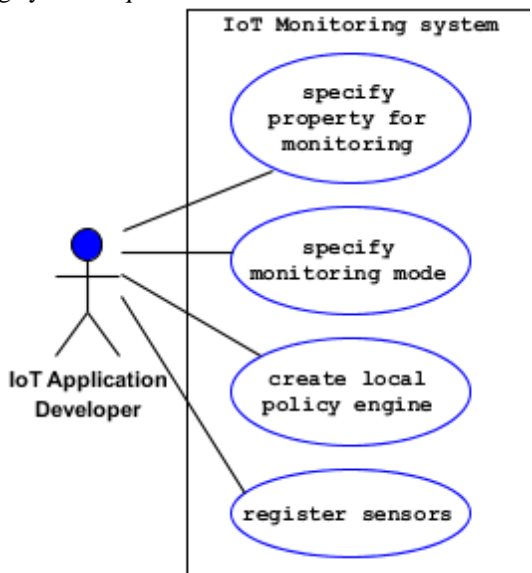

*Fig. 5. UML use case diagram for the monitoring system requirements.*

*Monitoring components:* these represent the main classes involved in the monitoring activity which are described as follows:

- *Sensor*: its sole responsibility is to collect data about the physical entity (thing) property that is of high importance to the adaptation process and then send it to the monitor. There are two kinds of sensor namely the time-triggered and event-triggered sensors.

- *System property*: Also referred to as the context element, this is the property that is of a direct connection and great interest to the adaptation process. This property is the target of the monitoring activity and the main concern of the monitor component is to keep its value within a desirable or acceptable range. Often, a threshold is used to accomplish this task. Examples of system properties include server load, server throughput, response time and bandwidth usage. The system property contributes to the runtime system state.

- *Environment property*: The environment property represents any contextual information that is external to the system in question and contributes to its runtime state. Examples of such properties include the time of operating, the current client connections in client-server architecture, etc.

- *Threshold*: This is the value that the monitor component will compare against to decide whether the current value of the system property is still within a desirable and acceptable range. An example of a threshold would be if room temperature becomes greater than

- *System runtime state*: At runtime, the system state is represented by the combination of the values of system properties and the properties representing the environment or the context within which the system is operating. Each system has a desirable state driven by its goals and non functional requirements. Often the deviation from this desirable state is the trigger of the adaptation process.

- *Monitor*: Its main tasks are to filter and aggregate data received from a set of sensors and send it to the *analyzer* (directly or indirectly) component for any further and usually complex analysis. In the IoT platform, big data analysis tools are typically used to handle the massive amounts of data generated by the physical entities. The aggregated data received from the sensors represent the system (or subsystem) runtime state at one particular point in time. In terms of software design patterns, the monitor and sensor are linked together with the observer design pattern as shown in Figure 6.

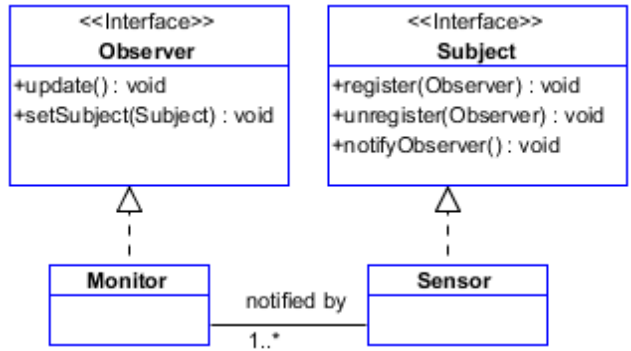

*Fig. 6. UML Class diagram for Monitor and Sensor relationship.*

- *Analysis system requirements*: The requirements of the analysis system in the IoT application are concerned with running big data analysis tools to extract some trends and patterns in the managed system behavior and accordingly issue a corrective action request. The corrective action could be either reactive or proactive. The former is a type of action taken in response to some undesirable situation which has already happened while the latter is acting based on predictions and anticipation of the future. However, the real-time nature of the IoT platform imposes the adoption of the proactive type where some algorithms and techniques (e.g. genetic algorithms) from the AI field are applied. The outcome of the analyzer component answers the question of whether an adaptation is required or not.

  Therefore, the following requirements are defined for the analysis activity:

  - Run data analysis tool
  - compose adaptation request

Figure 7 shows the UML use case diagram for the analysis system requirements.

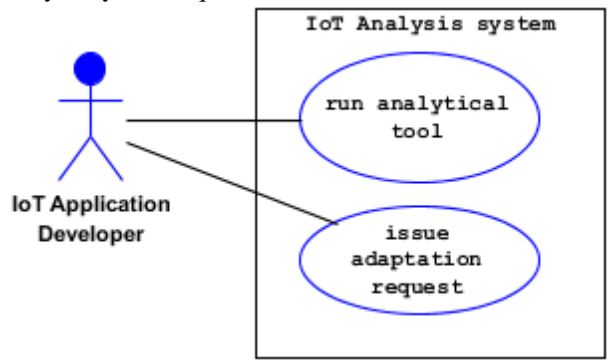

*Fig. 7. UML use case diagram for analysis system requirements.*

*Analysis components:* lists the main classes involved in the Analysis system and describes each class's responsibilities.

- *Analyzer*: Its responsibility is to receive logged data into the knowledge base (by monitor) and analyze them for any possible symptoms of system goals and requirements violation. The analyzer gets notified by the knowledge base component of the raising of new system state event. Therefore, it is linked with the knowledge base using the Observer design pattern where it plays the observer role and thus has to implement the observer interface. Once the analysis process has completed, the analyzer notifies the plan component of any necessary adaptations via sending an adaptation request.

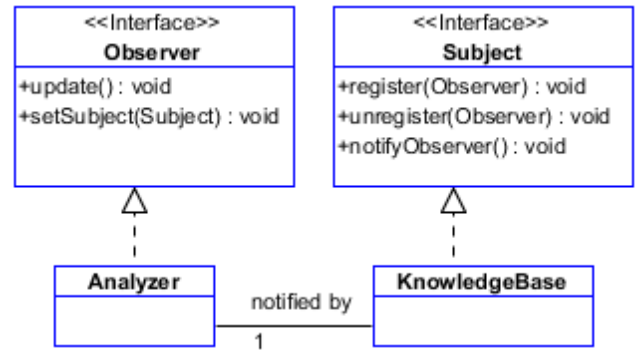

*Fig. 8. UML Class diagram for Analyzer and Knowledge relationship.*

- *Symptom*: Represents one of the undesirable states that the system in question must detect and take corrective actions against. A highly loaded server is an example of such symptoms. Symptoms work with a set of combined conditions and when these conditions are satisfied, the analyzer raises an adaptation request signal and sends it, along with the necessary information, to the plan component.

- *AdaptationRequest*: An adaptation request is created and sent to the plan component along with the necessary information. The latter includes the event describing the symptom (e.g. high patient temperature) and the frequency of this event in a specified time window (e.g. last hour).

- *SymptomRepository*: It contains a set of predefined symptoms that the system in question should avoid and heal up from once . It also provides a facility to add new emerging symptom at runtime via the *addSymptom* operation. This component is usually part of the knowledge base of the feedback control loop.

- *Planning system requirements*: The requirements of the planning system are concerned with constructing the change plan

which is composed of a set of corrective actions in response to an adaptation signal raised by the analyzer component. In this stage, the questions of what actions to be taken and in what order are answered. Therefore, the following requirements are defined for the planning activity:

- Compose change plan
- Dispatch change plan to execution system

Figure 9 shows the UML use case diagram for the planning system requirements.

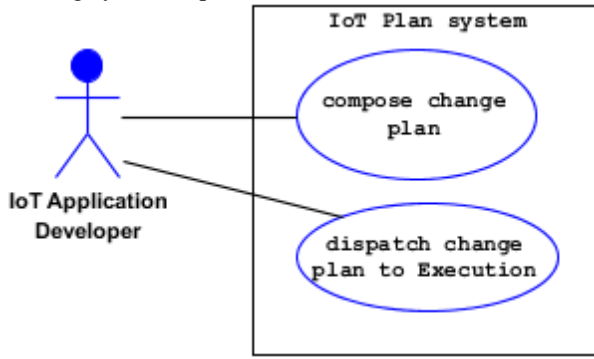

*Fig. 9. UML use case diagram for the planning system requirements.*

*Planning components:* lists the main classes involved in the planning system and describes each class's responsibilities as follows:

- *Plan*: It is responsible for constructing the change plan in response to an adaptation request received from the analyzer. The plan component uses the policy engine for accomplishing its task and then sends the constructed change plan to the execute component to dispatch these changes. The plan is linked with the analyzer using the observer design pattern where it takes on the observer role and thus implements the observer interface.

- *PolicyEngine*: It contains the policies (high level goals) that control the operating and functioning of the system in question. Policies may take the form of Event-Condition-Action (ECA) rules which determine the actions to be taken when an event is raised (or expected) provided some specific conditions are met. The policy engine belongs to the knowledge base of the feedback control loop. It provides the necessary interface for the system administrators to define and modify the policies of the system at hand.

- **ChangePlan.** It contains the actions that should be dispatched to the execute component in order to perform the adaptation and corrective actions. It is often called the strategy in which the actions are performed in specific and logical order.

- *Execution system requirements:* The requirements of the execution system are concerned with executing the adaptation actions or change plan that is received from the plan component. These actions must be executed in some specific order (sequentially or concurrently or maybe mixed of the two) as stated in the plan. The execution system uses a set of actuators to apply the required changes to the managed system which usually involve setting new values to the system or environment properties which are collectively constitute the system state. The following requirements are specified for the execution activity:

- Execute change plan
- Update system state

Figure 10 shows the UML use case diagram for the execution system requirements.

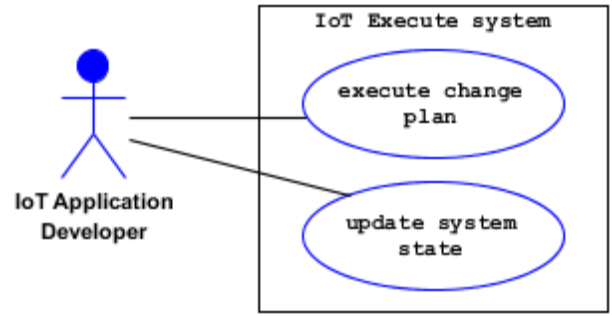

*Fig. 10. UML use case diagram for the execution system requirements.*

*Execution components:* lists the main classes involved in execution system and describes each class's responsibilities.

- *Executor*: It is responsible for sending the corrective actions to one or more effectors in a specific order.

- *Effector*: It is responsible for applying changes to system or environment properties according to some actions received from the executor component.

The central class of this activity is the executor which contains the update operation where it receives the change plan (corrective actions) from the plan. Once it has received the corrective actions, it dispatch them to a set of effectors to apply the changes to the target system and environment properties. Therefore, it is linked with the plan and effector components using the Observer design pattern where it plays both the observer role (with the plan) and the subject role (with the effector) and thus has to implement two interfaces, namely the observer and the subject.

- *Knowledge requirements*: The requirements of the knowledge system are concerned with the policy and event (or symptom) definitions and therefore the requirements here are as follows:
  - Define policy
  - Edit policy
  - Define event
  - Edit event
  - Log data or alerts

Figure 11 shows the UML use case diagram for the knowledge system requirements.

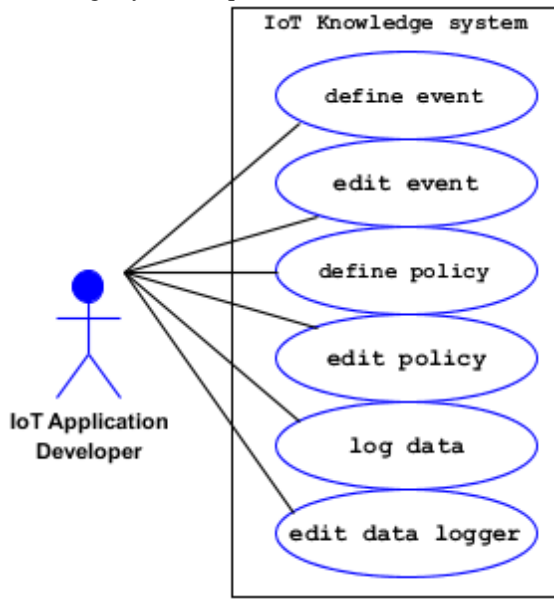

*Fig. 11.  UML use case for the Knowledge system requirements.*

- *Environment requirements*: The environment is defined as any external actor that affects the system in some way. Therefore, the environment property represents any contextual information that is external to the system in question and contributes to its runtime state. The environment requirement of the framework is as follows:
  - construct environment model.

Figure 12 shows the UML use case diagram for the environment system requirements.

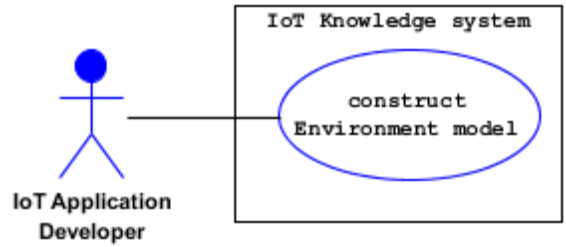

*Fig. 12.  UML use case for the environment system requirements.*

All of the requirements presented so far constitute the functions and capabilities that should be provided by the proposed framework. Such high level requirements are then detailed and expressed in terms of software programs using one of the programming languages.

## 4.4 Software Process For Iot Application Development

This section is dedicated to the software process or development methodology followed by our development framework. The proposed framework adopts the MDD paradigm to gain some valuable benefits which are very crucial in designing distributed systems in general and IoT applications in particular. Raising the abstraction level and separation of concerns are among those benefits. Achieving those design principles will result in a resilient system that can be a future proof and would survive in a world of rapidly changing system requirements and technologies, and full of heterogeneous devices, platforms and programming languages. Accordingly, our IoT application development methodology is divided into a number of fundamental stages: Platform Independent Model (PIM), Platform Specific Model (PSM) and the code. A description of these stages is presented below.

- Platform Independent Model (PIM)

The first stage of our software process is the Platform Independent Model (PIM) where the system under consideration is expressed in neutral concepts; no specific platform design decisions are made in this model. As stated in the proposed framework requirements section, the concepts used to model the system in question are Domain, Task, Service and Service composite. The artifacts produced in this stage are expressed in the form of XML documents.

- Autonomic Platform Independent Model (APIM)

In this stage, the software components that are responsible for handling the management and self adaptation aspects of IoT applications are defined. Components such as the monitor, analyzer and planner are specified here and associated with the physical devices (managed systems). However, these components are expressed here in a technology and platform independent manner in the form of XML documents.

- Autonomic Platform Specific Model (APSM)

At this stage, the specific elements and terms for a specific platform, Java web services for instance, are added to the model obtained in the previous stage. The resulted model is expressed in an XML document.

- Autonomic Code Generation

Generating autonomic code is performed at the last stage of the proposed process where the appropriate transformer is run for the autonomic code generation for a particular platform. Two Java based transformers are used here, one for generating the code for the core services and another to generate the autonomic components. To target Java Web services platform, for instance, the JavaCodeGenerator.java file is applied to the JavaWebServiceTemplate.java to generate the core Java web services. Likewise, the AutonomicJavaGenerator.java file is applied to the AutonomicJavaWSTemplate.java in order to generate the autonomic web services.

## 5. CONCLUSION AND FUTURE WORK

In this paper we have investigated the challenges that obstruct the realization of IoT as well as the promising solutions that pave the way for the acceptance and enabling of IoT. We focused specifically on the application layer and how to provide system developers with the right tools and methodology to develop IoT applications in a seamless and effective way. We believe that IoT applications are highly dynamic in nature and thus need to be engineered with the self adaptive and autonomic concepts in mind. Therefore, our proposed IoT software development lifecycle was based on the IBM architecture blueprint for autonomic systems. We also have taken the real-time nature of IoT applications and its influence on the organisation of the MAPE-K components into consideration. Accordingly, we adopted the master-slave pattern in which the adaptation logic takes a hierarchical relationship between one master component who is responsible for the analysis and planning of the adaptations and between a set of slave components whose responsibilities are to monitor properties of interest and execute the adaptation actions. To cater for the runtime dynamic and heterogeneity aspects of IoT applications, we adopt the MDD paradigm for our proposed development framework. Raising the software abstraction level, which is the central idea behind MDD, and postponing the adherence to a specific platform or programming language is a valuable requirement in the IoT platform.

A further work is required to address the following issues:

- More detailed design patterns for each component of the MAPE-K based IoT application development framework.
- A case study based evaluation of the proposed development framework.
- A development environment for modeling and simulating IoT applications.

## ACKNOWLEDGMENT

The author would like to thank his academic department for the precious support received throughout the work on this paper.

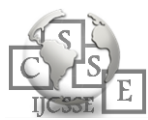